\documentstyle[12pt]{article}
\begin{document}
\newcommand{\lt}{\left}
\newcommand{\ity}{\infty}
\newcommand{\be}{\begin{equation}}
\newcommand{\ee}{\end{equation}}
\newcommand{\ri}{\right}
\newcommand{\fr}{\frac}

\baselineskip=12pt

In an earlier work,  several
properties of fundamental particles were brought together by a
simple equation based on continuity and discreteness. It is
shown here, that the maximum modes of decay of all fundamental
particles can also be predicted without any arbitrary
parameters. The method  used is  to break up the mean lifetimes
of particles  to obtain  the maximum modes of decay. This is
done  by using a binary expansion of  $\hbar/MT$ where M is the mass
of the particle and T is the mean lifetime. The agreements
between that obtained from theory and experiment are remarkable.
The ordering of the flavours  plays an important  part in
understanding the reasons for this agreement. It is shown that
the Zeno effect in Quantum mechanics is connected with use of
the binary series.
\\

In an earlier review$^1$  it was
shown that it was possible to systematise the mean values of
half lives of all fundamental particles, $\beta $ and $\alpha$
emitters and obtain the half lives of iso-spin  particles and in
particular the half-life of the proton from the half life of the
neutron which gives a value of 5.33 $\times$  10$^{33}$ years.
It was also possible to determine the parity of  $\beta$  emitters
and determine the energy levels of light nuclei. All this was
possible starting from a simple  nondimensional equation whose
relation to Quantum Mechanics is discussed below.  From the
equation
\baselineskip=28pt
 \begin{eqnarray}
  \frac{ \hbar}{MT} & = & \frac{n}{2^n}\cr    & = & \frac{\Gamma}{T}
\end{eqnarray}

\newpage
\baselineskip=24pt
where

$M$ is the mass of the concerned particle in MeV

$T$  is the mean decay half-life

$\hbar$   is Planck's constant$/2\pi$ and

$n$   is an integer
$$              \Gamma  \ \mbox{(the width)} \   =  \frac{\hbar}{M}
$$
it is possible to obtain  the values of  $n$  from known
measured values of $M$ and $T$.  In the case of fundamental
particles,  $M$ is the total mass of the decaying particle, as all
this mass is converted in the decay process.  In the case of
$\beta$  decay,  the concerned mass in Eqn. 1 is taken as the mass
of the neutron as in the Fermi theory.  In the case of $\alpha$
decay,  the mass is taken as the equivalent of the binding energy
of the nucleus that decays.

        We define a quantity $p  =  \log  2^n/n$ (log to base
10) .  It is an interesting  quantity in that, for certain
values of  $n$, $p$  becomes close to prime numbers and the
departures of the values of  $p$, i.e.($p-p_o$)  from the values
of the exact primes, give rise to partity in $\beta$ decay and the
flavours of fundamental particles$^2$.  The plot of log $Er/\Gamma$
where $Er$ is the energy levels of width $\Gamma$, against $Dn$,
which is the difference of the $n$'s obtained from Eqn. 1 from
the ground state to the excited state, suggests that the energies
transferred to the levels correspond to the masses of light
mesons.

        The  $n$-values for iso-spin singlets
and neutral members of the iso-spin triplets belonging to SU3
octets are found to be related to each other through some
periodicity.  The following equations show that $\pi^0\eta$  for
pseudoscalar mesonic octet, $\rho^0 \omega$ for vector mesonic
octet and $\Sigma^0,  \Lambda$ for  $\frac{1}{2}^+$octet are examples of
such pairs with iso-spin with $I_3$=0 which show the following
regularities:
$$ n_{\pi 0} - n = 2^2 + 1;  n_{\rho 0}
- n_{\omega} = 2^2 + 1;
n_f - n_a = 2^2+1; n_{\Lambda} - n_{\Sigma_{0}} = 2^5 + 1  $$
The others show a different kind of periodicity.  The two
different kinds of symmetries can be combined to give a more
general result through the relation
$$ \Delta n =  2^q + \Delta I $$
where $\Delta n$ and $\Delta I$  represent the differences in the
integer and iso-spin values.  The regularities found in the $n$
differences, point out to the quantisation of the $\Gamma/M$
ratios for the iso-spin levels,  which can be written as
 $$ \lt[\fr{MT}{\hbar} \ri]_{n+a} = k \lt[\fr{MT}{\hbar}\ri]_{n}
$$
where $n$ and  $\alpha$ are integers  i.e. $k = [n/(n+\alpha$)]
$2^{\alpha}$ which is true only when $n+a =2q$ for every $q \leq
 \alpha$.  The quantisation of iso-spins suggest $k$ should be an
integer.

        From this,  it can be shown that for $\Sigma^{\pm}$,  $n$
=54, $\alpha$=10,   $q$ = 6 but for  $\Sigma_0, \ q = 5, \ \alpha = 10,\  n =
22$  which matches with the value of $n$ for $\Sigma_0$.  A similar
quantification can be made to $\pi^{\pm}$ and $\pi^0$  and
several other iso-spin cases including that of the  proton.
All this is described in the review by Ramanna and Sharma$^1$
but a summary is given here for the sake of completeness.  The
figures in Appendix 1 give some more successes of the theory.

        The method for determining the
decay modes of fundamental particles is to consider a quantity
$\hbar/MT$  and expand it into a binary series. Each
dimensionless quantity,  $\hbar/MT$ which is always less than 1,
is converted to the sum of binary decimals.   The series comes
out  with  many zeros followed by a combination of 1's and 0's.
The   infinite number of zeros which follows the set between the
1's have  no particular significance.

        As shown earlier$^1$
$\hbar/MT$  can be written as a binary decimal as follows :
\be \fr{\hbar}{MT} =  \lt(a_1 \cdot 1 + a_2 \cdot \fr{1}{2} + a_3 \cdot \fr{1}{4} +
a_4   \cdot \fr{1}{8} \cdots + a_n  \cdot \fr{1}{2^n} \cdots  +
\ity\ri) \ee
where the $a$'s are either 0's or 1's.

                In this paper it is shown that the maximum modes
of decay can be obtained with only a knowledge of $M$ and $T$
through the binary splitting of   $\hbar/MT$. Table 1 (Col. C)
gives the value of  $n$  for all particles as derived from eqn.
1, from published data$^3$.

        The number of zeros, $H$ in the
binary expansion (Col. $H$) which precede  the 1's  is shown to
be  related to particle flavours$^1$ and is close to the value
of  $n$.  The combination of the 1's and 0's, converted to
ordinary decimals are  tabulated as Bin Dec (BD) and shown  in
Col.  $J$.

        We can  write e.g. for the $\pi^{\pm}$  meson
\begin{eqnarray}   \log  \fr{n}{2^n} &    = &  (.00 \cdots 53 Zeros)
 \cdot 2^{-54}  \cdot  \ \cdots)  \ \cdot \
(11101000 \cdots)   \cr
                        &  =  & 2^{-54} \lt(1+ \fr{1}{2}  + \fr{1}{4} + 0 + \fr{1}{16}
\cdots \ri)  \nonumber \end{eqnarray}
where  the number of zeros ($Z$)  is 53. It is shown below that the
expression
\be (H_m - H_{m-1}) \log 2  + \log \lt(\fr{BD_m}{BD_{m-1}}\ri) \ee
where $H_m$ is the number of zeros for the n$^{\rm th}$ meson arranged
according to flavours (Col. $B$), gives the maximum decay modes
(MDM) of the particles .

                Since   $p  =  n \log 2 - \log n$, the
sequential differences  can be written as
\be  (p_m   -  p_{m-1}) = \log 2 - \log  \fr{n_m}{n_{m-1}} \ee
which shows that it is similar in form with expression (3)

                                It was noticed purely
empirically, that the differences in (log $M)/BD$  from 1,  ordered
according to flavours, given in Col. $K$ are in good agreement
to  the measured MDM's. The consecutive absolute  differences
(Col. $N$) are  given by,
\be                                1 - \left[\fr{(\log M_m)}{(BD_m)} -
\fr{(\log M_{(m-1)})}{(BD_{(m-1)}}\right]  \ee
where $M$ is the mass of the particle in MeV and m is the serial
number  of the
particles ordered according to n's in flavours, i.e.  serial
nos. (S.No.) (Col. $B$).
The empirical Eqn. (5) is given here to show the dependence of
the MDM's  on mass differences.
\begin{itemize}
\item[Fig 1] is an  XY plots of Col. $O$ and  Col. $N$ which
relates exp$^n$ (3) and (5)
\item[Fig 2]   is  an XY plot of  MDM (Col. $D$)  and  Col. $N$
which relates the experimental values of MDM (exp$^n$) with
exp$^n$ (5)
\item[Fig 3]  is  an  XY plot  of MDM (Col. $D$)  and  Col. $O$
relates MDM with exp$^n$   (3)
\item[Fig 4]  is  an  XY plot  of  $p(\log (2^n/n))$  and  Col.
$O$ relates exp$^n$s (3) and  (4).
\end{itemize}

These figures  do not show any immediate order but they are,
however,  identical  as seen from  the LINE plot in fig 5a. The
rearrangement  takes place when the concerned values of n  are
shifted to agree with the appropriate $n$ ordering for that XY
plots. Figure 6 gives the same information as fig (5a) but shows
the possibility of predicting the MDM values of those particles
for which no measurement exists.  For those which have a
measured value two shades are seen, but for those which have no
measured values only one shade is seen corresponding to the
theoretical values (Col. $O$) (multiplied by 2 to bring out the
differences).  The predicted values of MDMs for these particles
are comparatively small, showing that it is difficult to measure
their MDM's.  Figure (5b) gives the predicted values of the MDM for
only those mesons whose values are not available.

        The agreement between the measured values and those
obtained from the empirical expression (4) equation as well as
that obtained from  the binary series is striking. In Fig 7
the XY plot  between  BD and number of zeros $Z$ is given, the numbers
along the points  are the corresponding values of n. They are
seen to be arranged according to favours.

By equating (3) and (5), one gets the equation
\begin{equation}
1 - \left[\log \lt(\fr{M_m}{BD_{m-1}}\ri) - \log \fr{M_{m-1}}{BD_{m-1}}
= \lt(H_m - H_{m+1}\ri) \log 2  - \log  \fr{BD_m}{\log BD_{m-1}}\right]
\end{equation}
showing that the difference of the log of the mass is empirically
connected to the binary expansion.

        The use of the binary series to
determine the break up of the mean life  can also be used to
connect it to the  Zeno effect in QM$^4$. Each entry of 1 or 0
can be looked as an interaction in time with the system and thus
acts like a mathematical clock whose time widths can be varied.

Writing (1) as  $1/T  =  M/\hbar (\Sigma)$ where $\Sigma$ is the
binary series, we connect it with the probability of survival of
a decaying system $p_N$. After an $N$  number  of observations in a
finite time $T$ at intervals of $\Delta t$ we can write that
$$
P_N = \lt[1-\lt(\fr{C}{4N^2}\ri)\ri]^N $$
where $C$ is a constant and $N  = T\Delta t$.

                We also know that {\bf time} is a continuous
element and we define its continuous nature as was done by
Cantor$^5$.  The Cantor theorem which is of great generality
states that given a set $A$, it is possible to construct another
set $B$ with a greater   cardinal number. If $A$ consists of
three numbers, then $B$ consists of 2$^3$ numbers and in general
for every $N$ there is  associated a set of values which
corresponds   to the continuous  elements i.e. 2$^N$, a fact
used in the formulation of eqn. (1).  The Cantor theorem further
states that if a set contains $N$ elements where the $N$'s are
positive integers, then there exists a set  $B$ with $2^N$
elements and if $A$ consists of all positive integers, $B$ is
equivalent to the continuum of real numbers from 0 to 1.  Thus
for a continuum of time which is always continuous even in
finite intervals,  $2^N$ has to be used.

If  the $N$'s above is replaced by $2^N$ we get the probability
for survival for continuous observation and the ratios of the
survival probabilities for discrete and continuous parameters
can be written as
 \begin{eqnarray}
\frac{\log \ p_{\rm sur}^{\rm
discrete}}{log \ p_{\rm sur}^{\rm continuous}}  & = &
\frac{N \log
[1-C/4N^2]}{2^N \ \log  [1- C/4 \cdot (2^N)^2]}\cr
& = &  \fr{N}{2^N}  \cdot \alpha   \end{eqnarray}

 To a first order we can write the nonsurvival probability as                                                                     pNON$_{\rm sur}$   =  log (1 $-$ pNON$_{\rm sur}$) =log p$_{\rm sur}$
The ratio of the {\bf non-survival} probabilities, if $\alpha = 1$,
is now  given by  $N/2^N$ and this is identical to the ratio
$\Gamma/T$  given   in  Eqn. (1), i.e. the width divided by the mean
life time. As is known the width of a spectral line is
continuous in nature, whereas the energy is discrete.  It is
remarkable that such a simple equation as (1) is able to bring
to order so many facts about  fundamental particles and some
allied aspects of nuclear physics. In a sense n and 2$^n$ act as
"measures" of discreteness and continuity and  the  results
obtained show, that they are well within the frame work of
Quantum Mechanics.

        My thanks are due to Prof. B.V Sreekantan for his
continuous support and many useful discussions. I am grateful to
Prof. Ramachandra for his comments on number theory and leading
me to the binary decimal series. My thanks are due to Dr.
Dipankar Home for his constant encouragement and his introducing
me to the theory of the Zeno Quantum  effect. I must thank
Prof. Asoke Mitra for his valuable comments at various stages of
the work. I am also thankful to Prof. C.V. Sundaram for many
useful discussions.  My grateful thanks are due to Mr K.S Rama
Krishna and Mr. Manish Chauhan for their  technical help. \\

\noindent {\bf References}
\begin{enumerate}
\item Raja Ramanna and Anju Sharma {\it Current Science}, 1997, {\bf 73},
pp.1083-1097.
\item  Raja Ramanna,  {\it Current Science},   1998, {\bf 75}, pp.183-186.
\item  Review of Particle Properties, {\it Physical Review},
1994,  {\bf D50},  Part I.
\item  Dipankar Home,  {\it Conceptual Foundations of Quantum
Mechanics-An overview from modern perspectives},  Plenum
Press,  1998.
\item Courant,  R and Robbins, H., {\it What is Mathematics: An Elementary
Approach to Ideas and Methods}, Oxford University Press, New
York,  1978, pp.84
\end{enumerate}

\end{document}